\documentclass[11pt]{article}
\usepackage{geometry}                % See geometry.pdf to learn the layout options. There are lots.
\usepackage[parfill]{parskip}    % Activate to begin paragraphs with an empty line rather than an indent
\usepackage{graphicx}
\usepackage{subfigure}
\usepackage{amssymb}
\usepackage{amsmath}
\usepackage{multirow}
\usepackage{cite}
\usepackage{setspace}
\title{
Physical Model of the Immune Response of Bacteria Against
Bacteriophage Through the Adaptive CRISPR-Cas Immune System }
\author{Pu Han$^1$, Liang Ren Niestemski$^1$, Jeffrey E. Barrick$^2$, and Michael W. Deem$^1$ \\
\hbox{}$^1$Department of Physics \& Astronomy\\
Rice University\\
Houston, TX  77005 \\
\hbox{}$^2$Department of Chemistry \& Biochemistry\\
The University of Texas at Austin\\
Austin, TX 78712 \\ }

\begin{document}
\maketitle

\begin{abstract}
Bacteria and archaea have evolved an adaptive, heritable immune system
that recognizes and protects against viruses or plasmids. This
system, known as the CRISPR-Cas system, allows the host to
recognize and incorporate short foreign DNA or RNA
 sequences, called `spacers' into its CRISPR system. Spacers
in the CRISPR system provide a record of the history of bacteria and
phage coevolution. We use a physical model to
study the dynamics of this coevolution as it evolves stochastically
over time. We focus on the impact of
mutation and recombination on bacteria and phage evolution and
evasion. We discuss the effect of different spacer deletion mechanisms
on the coevolutionary dynamics.  We make predictions about bacteria and phage
population growth, spacer diversity within the
CRISPR locus, and spacer protection against the phage population.
\end{abstract}

\section{Introduction to CRISPR}

\subsection{CRISPR}

\begin{figure}
%\centering
\includegraphics[scale=0.4]{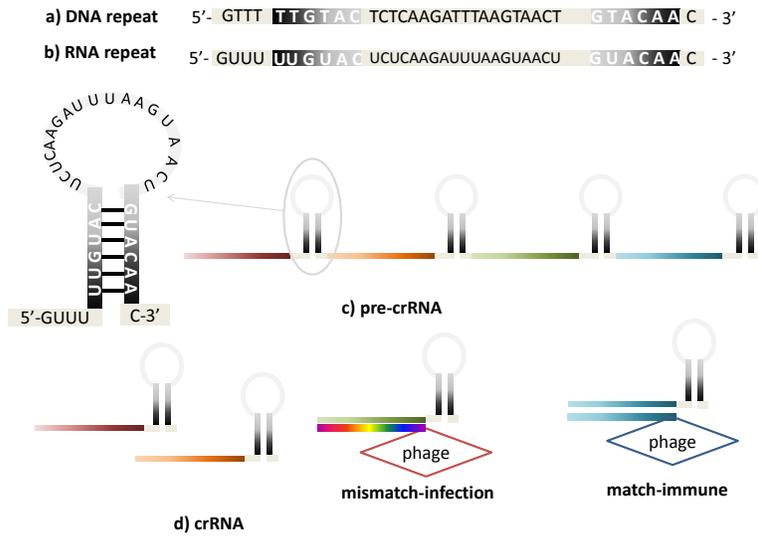}
\caption{a) Typical repeat in \emph{S.\
thermophilus} CRISPR1, data taken from \cite{HorvathJB2008}. b) RNA
repeat. c) Left: secondary structure (hairpin) of RNA repeat.
right: pre crRNA. d) Targeting and protecting during CRISPR
immunity. \label{fig:crispr}}
\end{figure}

The newly discovered CRISPR system in bacteria and archaea
is a fascinating system for experimentalists and
theorists. CRISPR was initially discovered in the gene sequence of
\emph{Escherichia coli} \cite{IshinoJB1987}. Ishino et al.\ found an unusual
structure in the 3'-end flanking region of the \emph{iap} gene, namely
repeats of the same 29 nucleotides, each separated by spacers of 32 non-repeatable
nucleotides. Inside each repeat there are two short
sequences of DNA that are nearly reverse complements of each other,
\emph{i.e.}\ nearly palindromic sequence. These two palindromic
DNA sequences, \emph{e.g.}\ TTGTAC and GTACAA in Fig.\
\ref{fig:crispr}a, are transcribed into RNA
sequences \emph{e.g.}\ UUGUAC and GUACAA in Fig.\
\ref{fig:crispr}b, that can base pair to form a stable hair pin loop, as in Fig.\
\ref{fig:crispr}c. The discovery of these repeat sequences in \emph{E. coli} spawned an
extensive search for similar interspersed and repetitive DNA
sequences in other bacteria and archaea. To date these structures
have been identified in 40\% of bacteria and 90\% of archaea
\cite{GrissaNAR2007,HovathBarrangouScience2010,SorekNatureMicro2008,OostTBS2009}.
They are now termed clustered regularly interspaced short
palindromic repeats (CRISPR) \cite{JansenMolMic2002}.

Bacteria or
archaea can encode one or more CRISPR systems in their genome.
Although the CRISPR gene structure varies greatly between
different species, it has a few common features. It always has a
leader-repeat-spacer-repeat-spacer{\ldots}\ organization. Repeats are the
regions with the same nucleotide sequences with nearly palindromic
symmetry. The length of the repeat ranges from 23 bp to 47 bp in different organisms.
Spacers are the nucleotide regions between the repeats.
The length of the spacer ranges from 21 bp to 72
bp in different organisms \cite{HovathBarrangouScience2010}. Leaders are AT rich sequences
located at the 5' end of the CRISPR system. Leaders serve as an
indicator of the beginning of the CRISPR system and give it a polarity. Leader sequences
of different CRISPR systems in the same species are the
same \cite{JansenMolMic2002,HorvathJB2008}. Leaders serve as
the recognition site for the addition of new spacers, and new
spacers are always added to the leader proximal end of the
CRISPR \cite{BarrangouScience2007,DeveauJB2008}.

\subsection{CRISPR is an immune system}

CRISPR is part of the immune system of bacterial and archaea. This
functionality was discovered while studying phage resistance in
\emph{Streptococcus thermophilus}, a lactic acid bacterium used in
the production of yogurt from milk
\cite{Sturinonrmicro2006,BrussowARM2001,BrussowAvL2002}.
Like other types of bacteria, milk lactic acid bacteria can be
infected by viruses known as bacteriophage, and phage infection is the major cause of milk
fermentation failure. It was observed that not all of the milk
lactic acid bacteria cease to grow upon challenge with
bacteriophage \cite{DeveauJB2008,HorvathIJoFM2009}. Some of
the bacteria were phage resistant. Extensive genome sequencing of lactic
acid bacteria and virulent phage led to a better understanding of
the difference between phage resistant bacteria and phage
sensitive bacteria. Bacteria with phage resistance have CRISPR systems in
their genome with spacer sequences that match the specific phage
to which they are resistant. Upon challenge with a new type of phage,
vulnerable bacteria strains have the ability to acquire
sequences, termed protospacers, from the phage genome that are inserted into their
CRISPR next to the leader sequence. This newly acquired spacer
contains genetic information from the population of currently infecting phage.
Descendants of these bacteria inherit their ancestor's genome with
the inserted spacers and are phage resistant. The connection
between the CRISPR related immune system, and phage resistance was
confirmed in several laboratory experiments
\cite{BarrangouScience2007}. A phage-resistant strain of
bacteria remained phage resistant upon removal of all spacers
except the one derived from the phage of interest. A phage resistant
strain became phage sensitive upon removal of the specific spacer
derived from the phage of interest even when all the other
non-relevant spacers were present. When the matching spacer was added back into the
CRISPR, the previously phage-sensitive bacteria became resistant to
that specific phage, but were
susceptible to new phages with different genomes. Bacteria can be
immune to more than one type of phage if different CRISPR
spacers match different phage genomes. To sum up, a CRISPR spacer matching a protospacer sequence in a phage genome
provides resistance against that specific phage.

\subsection{Mechanism of CRISPR action}

The CRISPR system as well as the CRISPR associated (Cas) genes located
in the vicinity of CRISPR are essential components of a
functional CRISPR-associated complex of antiviral defense
(Cascade) system. The components of the Cascade system can be isolated from bacteria and reconstituted for study in vitro,
and the mechanism of Cascade action has been investigated in
both \emph{E. coli} and \emph{S.
thermophilus} \cite{BrounsScience2008,BrounsNaturesmb2011,BarrangouScience2007,GarneauNature2010}.
The Cascade defense process starts with spacer acquisition,
proceeds with CRISPR expression, and finishes with CRISPR
interference. During spacer acquisition, Cascade recognizes a
foreign nucleic acid sequence, \emph{i.e.}\ a protospacer. In the case of
\emph{S. thermophilus}, short conserved regions within a few bases
of the protospacer sequence are identified as CRISPR motifs. These
CRISPR motifs serve as signal for the bacteria's Cascade system to
recognize the protospacer. Upon recognition, a new sequence of
nucleic acid identical to the protospacer is generated and
integrated into the CRISPR system as a newly acquired spacer.
During CRISPR expression, the spacer is transcribed into
pre-crRNA. With the participation of the Cas protein, pre-crRNA
matures into small crRNAs \cite{BrounsScience2008}. Within each
crRNA, the transcript of a single palindromic repeat folds into a
stable hairpin  shape termed a handle with several stable base
pairs, which may serve as a platform for RNA-binding Cas
proteins \cite{HorvathJB2008}. The CRISPR spacer is
connected to one end of the hairpin structure in each crRNA transcript. The crRNA
is transported to the target phage DNA
\cite{ShahTransaction2009,MarraffiniScience2008,MarraffininatRevGenet2010} or
RNA \cite{HaleCell2009}. During CRISPR expression, the crRNA
guide the Cas complex to foreign nucleic acids. The expressed
spacer sequence provided by the crRNA is thought to recognize and
guide the complex to bind the specific protospacer target sequence.
Cas proteins with nuclease activity then cleave the invading nucleic
acids in order to inhibit phage infection.
If there is no match between the CRISPR
spacer and the phage DNA, the phage is not neutralized by
the spacer transcript. In this case, the phage can reproduce inside the bacteria
and lead to bacteria lysis and death. This process is illustrated in Fig.\
\ref{fig:crispr}d.

\subsection{CRISPR maintenance}

\begin{figure}
%\centering
\includegraphics[scale=0.5]{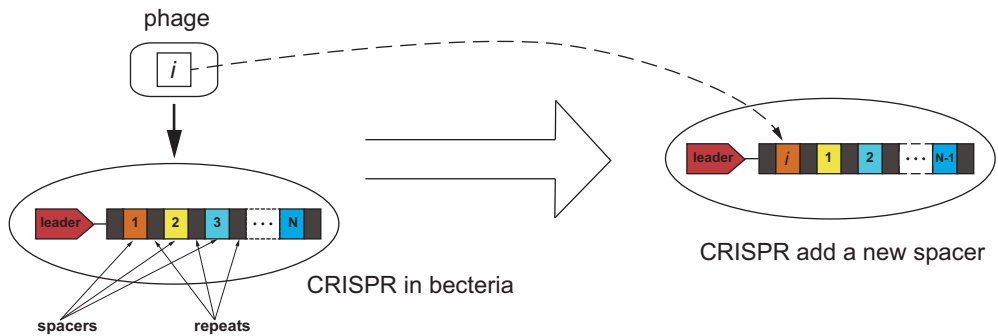}
\caption{Addition of a new spacer to the CRISPR locus at the
leader-proximal end. The protospacer in virus $i$ added to CRISPR. Spacers $1$--
$(N-1)$ are shifted to the
 right. Spacer $N$ is deleted. Other deletion mechanisms will
be discussed later. \label{fig:addanewspacer}}
\end{figure}

CRISPR can acquire new spacers from protospacer sequences within a  phage genome.
This process is shown in Fig \ref{fig:addanewspacer}. Although the
addition of a single new spacer is a low probability event, it
can occur in  at least some cells within a population
 of bacteria upon the phage challenge.
Newly acquired spacers are inserted at the leader-proximal end
\cite{HorvathJB2008,GarneauNature2010}. The number of
repeat-spacer units per CRISPR ranges from a few to
hundreds \cite{MarraffininatRevGenet2010,DeveauAnnurevmicro},
with a typical length of 30--100 spacers. For
example, one strain of  \emph{S. thermophilus} has
32 spacers \cite{DeveauJB2008}. Since the CRISPR cannot grow to
infinite length, deletion of older spacers is required, and
deletion has been documented concomitantly with spacer addition.
The mechanism of spacer deletion is unclear, especially the
location of deletion. One hypothesis is that the oldest spacer is
the least needed in the current viral environment and should be
deleted. Under this hypothesis, the deletion always
happens at the leader distal end of the CRISPR. Another hypothesis is
that the deletions can happen in the middle of CRISPR locus at
random locations, or at locations following a certain
distribution, such as linear distribution within the cluster, perhaps by
recombination. Deletion at the leader-distal end and internal
deletion are found in bacteria \cite{GudbergsdottirMMi2011}.
Moreover, deletion of greater than one spacer at a time is also
observed in \emph{S. thermophilus} \cite{DeveauJB2008}. Taken altogether, these
experiments indicate that maintenance of CRISPR system by spacer
addition and deletion occurs.

\subsection{Bacteria/phage co-evolution}

The CRISPR immune system imposes a selection pressure on the
phages. Conversely, the phages also impose a selection pressure on
the bacteria. The efficiency of the CRISPR immune system has a
direct impact on the fitness of the bacteria. Since bacteria are
surrounded by multiple strains of phages, bacteria with CRISPR-containing
 spacers matching many phages are more likely to survive
and reproduce. Due to evolution of the phages, the CRISPR spacers must be
continually updated to protect against new phage sequences. The phage fitness
depends on its ability to avoid recognition by CRISPR. There are
several mechanisms of phage evolution. Phages can overcome CRISPR recognition
by acquiring a single
mutation \cite{BarrangouScience2007,DeveauJB2008}. This will
cause a mismatch between the spacer transcript of crRNA and the
invading phage protospacer, leading to failure of the CRISPR
interference. Another CRISPR-evading strategy is recombination
between phages during coinfection in bacteria.
 Recombination is an inherent feature of phage
evolution. Metagenomic studies of different phage population
document large scale recombination in phage
\cite{AnderssonScience2008}. Recombination can lead
to a more rapid rate of phage evolution away from CRISPR
recognition than does point
mutation.
First, recombination of previously established mutations
incorporates mutations that have already been selected for increased fitness, {\emph
i.e.}\ mutations at less risk of altering essential protein function. Second,
recombination can integrate multiple beneficial point mutations
in one step, and it may be the case that $l > 1$ mismatches
between the crRNA and protospacer are required for the phage to
escape CRISPR recognition. Thus, we expect recombination will allow
phage to evade CRISPR more effectively than point mutation
alone.

\subsection{Motivation}

The role of  recombination has been under-studied, even though it
is a significant driver of
evolution \cite{MarraffininatRevGenet2010}. Recombination and other
mechanisms for generating genomic diversity are especially
important in coevolving systems with large population density. Here,
we study the effect of recombination on bacteria-phage
coevolution. We use a physical model that incorporates selection
pressure, and we allow both recombination and mutation to occur.
The theory and model provide time-resolved `snapshots' of this
coexistence. We focus on the case where bacteria and
phage coexist, \emph{i.e.}\ neither bacteria nor phage are driven
extinct.  That is, we choose parameters of the system to establish
a robust coexistance so that bacteria and phage both have stable
populations without species extinction. We study the effect of
different spacer deletion mechanisms, which is important for
understanding how CRISPR functions and interpreting patterns of CRISPR variation in natural populations of bacteria.

In this paper we establish a physical model of bacteria-phage
convolution and study the impact of recombination, spacer
deletion, and non-linear growth dynamics in this system. In the
Methods section, we introduce mean field population dynamics
equations and a stochastic simulation to sample the underlying
Markov process. In the Results section, we discuss the effect of a
non-linear density-independent growth rate on the population
dynamics. We demonstrate heterogeneity of spacer diversity in
different spacer locations within CRISPR. We predict the spacer
usage with respect to spacer location. We investigate three
different deletion mechanisms: deletion of the oldest spacer,
deletion of older spacers with increased probability, and deletion
of a random spacer. We study the impact of recombination and
mutation on the evolutionary dynamics. In the Discussion session,
we relate observations from our
physical model to experimental and natural coevolution of bacteria and phage.
We conclude in the last section.

\section{Methods}
\subsection{Co-evolution model}

We consider a coexisting  system composed of bacteria and phages.
Each bacterium can have a different CRISPR system and there are multiple strains of phages. The
evolution of the bacteria and phages is interrelated and changes dynamically. Bacteria with higher fitness have more
descendants, and the number of bacteria with that CRISPR system increases.
At the same time, phages with unsuccessful CRISPR evading strategies
cease to reproduce, and the number of those strains
decreases. Since there are different sequences of bacteria and strains of phages,
the total population of both the bacteria and phages can reach a
steady state even though the population of each bacterial sequence or phage strain may be
changing with time. This steady state is what we are interested
in. The total number of bacteria and the total number of
phages reach the maximum steady-state values, after an initial
exponential growth phase.

\subsection{Events}

We describe the bacteria-phage community dynamics using a
population dynamics model \cite{DeemPRL2010}.  The population
structure of the bacteria and phages changes based on several
events. The bacteria can reproduce at a certain rate until they
reach the maximum capacity, defined by the available resources. This
rate can be constant or dependent on the phage population. The phages
reproduce at another rate, which can  also be constant or dependent
on the total number of bacteria, until the maximum carrying capacity
of the phages is reached. Upon exposure to phage, a bacterium has
the opportunity to  acquire a protospacer from that phage, which
will be inserted into the leader-proximal end of CRISPR in that
bacterium.
%If this occurs then the phage is neutralized and no longer reproduces.
 We assume that phages can mutate at some
defined rate or they can recombine with other phages, also at a
certain rate. Either process leads to avoidance of CRISPR
recognition by the evolved phages. Recombination has the additional
advantage that it can combine the benefits of multiple
mutations, which can provide the recombined phage with a higher
fitness.

\subsection{CRISPR details}

The number of spacers that a CRISPR contains varies between types
of bacteria. Most CRISPR contain fewer than 50 spacer repeats.
We here set the maximum number of spacers in CRISPR to
be 30. Upon phage attack, a new spacer can be acquired and
inserted at the leader-proximal end of CRISPR. We label the
leader-proximal position of the spacer to be position 1, and the
leader-distal position of spacer to be position 30. In general,
positions with smaller index host ``younger" spacers. When a new
spacer is added to a CRISPR that already has 30 spacers, spacer
deletion occurs to maintain a maximum length of CRISPR. We
investigate three different types of spacer deletion: delete the
``oldest" spacer; delete a spacer with a possibility proportional
to its distance to the leader end, and delete spacer at a
random position.

\subsection{Phage details}

Each phage genome is assumed to contain only a single protospacer.
This protospacer is what our dynamics depend on, so we track only the
protospacer part of the phage genome.
It is known that a single phage may contain multiple 
protospacers \cite{BarrangouScience2007},
often  localized in the early expressed, coding region of the phage genome 
\cite{DeveauJB2008}.  We here simplify the biology, assuming only
a single protospacer per phage.
The protospacer in the
genome of each phage is expressed as a bit string. Each bit of
the string can be either ``0" or ``1". The length of the phage bit
strings is $n$, and there are $2^n$ types of phage genomes. In
our simulation, we set $n = 10$. Therefore, we have $2^{10}$
genome types available for phage. Initially, the population
distribution of phage follows a logarithmic distribution $p(i) =
\log(150) - \log(i), i = 1,\ldots,150$, where $p(i)$ is proportional to the
percentage of $i^{th}$ phage strain. This distribution has been
used to fit experimental data \cite{TysonEM2007}. We start with
149 strains of phage with this distribution, and they evolve over
time.

When phage replicate, there is a chance for phage to mutate with a
rate $\mu$ per genome per replication. This is part of the CRISPR-evading strategy of phage.
We choose a random location in the phage genome
to be the location of the point mutation. As this location, we
alter the phage sequence from ``1" to ``0" or from ``0" to ``1".
The probability for $n$ mutations in one sequence is $\mu^n e^{- \mu}/n!$.

\subsection{Mean field approximation and Monte Carlo method}

We used two methods to study this system: a standard numerical
fourth-order Runge-Kutta method to solve the mean-field differential
equations and a stochastic simulation using the Lebowitz-Gillespie
algorithm \cite{GillespiePhysChem1977} to sample the Markov
process. Both methods converge to the same result in the limit of
a large population.

In the mean-field or infinite-population, spatially homogeneous
limit the system can be described by the differential equations \cite{DeemPRL2010}
\begin{eqnarray}
\frac{d x_{i,j}}{d t} &=& \left[c x_{i,j} - \beta \sum_{k \neq i,j} v_k
x_{i,j} + \beta \gamma \sum_m x_{j,m} v_i - \beta \gamma \sum_k
x_{i,j} v_k\right]\Theta\left( x_{max} - \sum_{i,j} x_{i,j} \right)\label{1} \\
\frac{d v_k}{d t} &=& \left[r v_k - \beta
\sum_{i,j} x_{i,j} v_k (\delta_{i,k}+ \delta_{j,k})\right]\Theta\left(v_{max} - \sum_k v_k \right)\label{2}
\end{eqnarray}
We have set the number of spacers in CRISPR to 2 initially. We also initially do not consider 
virus evolution.  The
population of bacteria with spacer $i$ in position 1 and spacer $j$ in position 2  is $x_{i,j}$, where the maximum bacteria
population is $x_{\rm max}$. The phage population is $v_k$, where the
maximum phage population is $v_{\rm max}$. Here $\Theta\left( x_{\rm max} -
\sum_{i,j} x_{i,j} \right)$ is a step function. When $ x_{\rm max} >
\sum_{i,j} x_{i,j} $, $\Theta$ has a value of 1, otherwise it is 0. Each
population grows until it reaches its maximum value.  
Maximum population sizes are given from ecology and are due to maximal carrying capacities
in the case of bacteria or number of available hosts in the case of viruses.
The dynamics
of the population depends on the events described earlier.
Bacteria grows at a rate $c$ until they reach the maximum
population. Phage grow at a rate $r$, which could be dependent or
independent of bacteria population $\sum_{i,j}x_{i,j}$. Bacteria have
an exposure rate $\beta$ to the phage. Upon phage attack, bacteria
have a probability $\gamma$ of acquiring a new spacer from the
protospacer in the phage genome, independent of the current
spacers within the CRISPR. The rate of spacer addition is
$\beta \gamma \sum_k v_k$.
Conversely, when the bacterial CRISPR system is
unable to recognize the invading foreign genetic material, lysis
of the bacteria occurs after infection. The rate of bacteria
killed by phage is $\beta \sum_{k \neq i,j} v_k$. The term $
\beta \gamma \sum_m x_{j,m} v_i $ represents the process of
converting other types of bacteria into type $i,j$. The term $
\beta \gamma \sum_k x_{i,j} v_k$ represents the process of
converting type $i,j$ into other types of bacteria.
 When the CRISPR locus contains spacers matching the viral
 genetic profile, the phage is disarmed
and eliminated. The rate of phage killed by bacteria is $\beta
       \sum_{i,j}x_{i,j}(\delta_{i,k}+\delta_{j,k})$.

\begin{figure}
\centering
\includegraphics[scale=0.5]{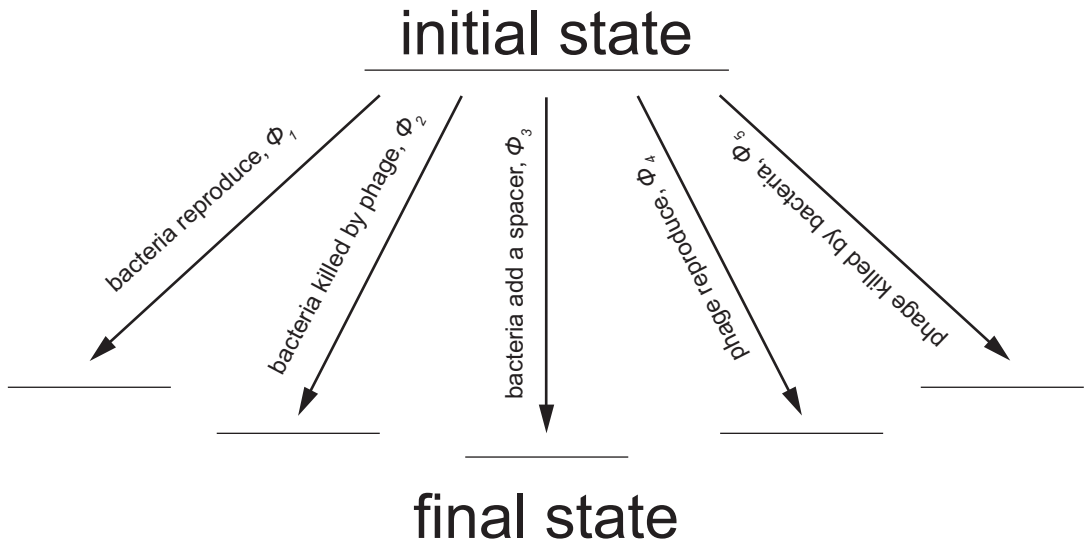}
\caption{In the Markov process, five categories of transition events change the state of
the system. The rates of all of these events are denoted by $\phi_i$.
Processes 2, 3, and 5 all result from phages infecting the bacteria.
Processes 4 results from phage infecting some bacteria, which could be the
population under study, or a different host population of bacteria.  There is
an additional category of events, not shown in this figure, which is
evolution of the virus due to mutation or recombination.
 \label{fig:LG}}
\end{figure}

We also use the Lebowitz-Gillespie algorithm to sample the
stochastic process of bacteria phage coevolution.  In this stochastic
process we include the mutation and recombination events
described in Section 2.4.   The
Lebowitz-Gillespie algorithm computes trajectories for
a Markov process in which the rate $\phi_i$ of the every event $i$ is known.
The algorithm works as follows:
% this is the description of the LG algorithm, do you think it is clear?
at time $t=0$ a list of all possible rates $\phi_i$ in the
system is formed. One event is randomly chosen to happen
from the list with a probability proportional to its rate. There are five
categories of events in the list. 1)  A bacteria can be randomly
chosen to reproduce at a rate $c$. Overall, bacteria reproduce at
a rate of $\phi_1 = c  x$, where $x=\sum_{i,j}x_{i,j}$ is the
total bacteria population and $x_{i,j}$ is the population of
bacteria strain with spacers $i$ and $j$. 2) A bacteria can be killed
by phage with a rate of $\beta \sum_{k \neq i,j} v_k$, where $v_k$
is the population of phage strain with protospacer $k$. Overall,
bacteria are killed by phage at a rate of $\phi_2 = \sum_{i,j}
\beta \sum_{k \neq i,j} v_k x_{i,j}$. 3) A new spacer can be added
to a randomly chosen bacteria with a rate of $\beta \gamma v$,
where $v = \sum_k v_k$ is the total phage population. This new
spacer is chosen from the protospacers among all the phage
according to the rate $\beta\gamma v_k$. Overall, new spacers can
be added to bacteria with the rate $\phi_3 = \beta \gamma v x$. 4)
A phage can be randomly chosen to reproduce at a rate of $r_0$.
Overall, phage reproduce at a rate of $\phi_4 = r v$. 5) A
phage can be killed by bacteria at a rate of $\beta
\sum_{i,j}x_{i,j}(\delta_{i,k}+\delta_{j,k})$. Overall, phage are
killed by bacteria at a rate of  $\phi_5 = \sum_k \beta
\sum_{i,j} v_k x_{i,j}(\delta_{i,k}+\delta_{j,k})$.  Time is
incremented by $-\ln(u)/\sum\phi_i$, where $u$ is a uniform random
number $\in (0,1]$. The rates of all possible events are then
updated, if they have changed. We iterate this process until time reaches
the specified final time. See Fig.\
\ref{fig:LG}. When the maximum population size is reached and a growth move is
attempted, a random
member of the population is deleted during replication.

Initially, we start with 149 types of phage with a logarithmic-decay
population structure, $p(i) = \log(150) - \log(i), i = 1,\ldots,150$, where
$p(i)$ is proportional to the percentage of $i^{th}$ phage strain.
%See Fig.\ \ref{fig:initphage}.
Every new
bacteria has a CRISPR with 30 empty spacers, \emph{i.e.}\ all value of
spacers are null. The initial population of phage is 1000,
the initial population of bacteria is 4000.

%\begin{figure}
%\centering
%\includegraphics[width=100mm]{virus.eps}
%\caption{The initial distribution of phage strains.
%\label{fig:initphage}}
%\end{figure}

\subsection{Density dependence of growth rate}

\begin{figure} \centering
\includegraphics[width=130mm]{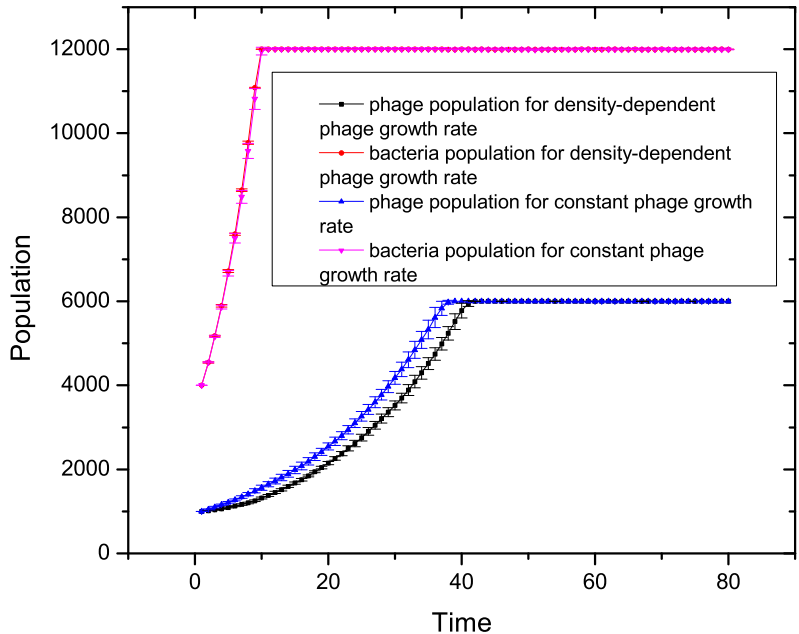}
\caption{Population of bacteria and phages with
time. We show results for constant and density-dependent phage growth rates.
The parameters are $c = 0.15, r = 0.05, \beta = 2 \times 10^{-5}$, and $
\gamma = 0.1$. The mutation rate per sequence per replication is $\mu = 0.01$. The maximum
population of phage is $v_{\rm max}=6000$, and the maximum population of bacteria is
$x_{\rm max}=12000$. The maximum number of spacers in a CRISPR is 30.
When the number of spacers in the CRISPR
array is over 30, the oldest spacer is deleted from the leader-distal end.
 There are 149 phage strains with a logarithmic initial population
distribution.
%, shown in Fig. \ref{fig:initphage}.
 \label{fig:population}}
\end{figure}

The model of the previous section applies when the phage grow
not only in the bacteria we study, but also in another set of
background bacteria.  These background bacteria are
the hosts providing
the approximately constant growth rate of the phage, $r_0$.
The populations of bacteria and phage are dynamically changing
with time. At short times, starting from an initially small
population, the bacteria grow exponentially until stabilizing at
the maximum population size. This is shown by the magenta curve
overlaid on top of the red curve in Fig.\ \ref{fig:population}.
We now set the maximum number of spacers to be 30.
Similarly, the phage population grows exponentially for a short
period of time until stabilizing at the maximum phage population size.
If the background bacteria are quite numerous, then
the phage can have a growth rate
independent of the bacteria under study,  labeled by $x$.

If there is no such set of background bacterial hosts,
the phage growth rate may depend directly on the bacteria we study,  labeled by $x$.
In this case, the reproduction rate of phage $k$ is a time dependent
function of the bacteria population, \emph{i.e.}\ $r_k = r_0
\sum_{i \ne k,j \ne k}x_{i,j}/ x_{\rm max}$,
where $x_{\rm max}$ is the maximum population of bacteria.
The average replication rate is
$\langle r_k \rangle =  \sum_k \sum_{i \ne k,j \ne k}x_{i,j} v_k/ (x_{\rm max} \sum_l v_l)$.
A simplified form if most of the bacteria population is available to any given phage is
$r = r_0
\sum_{i,j}x_{i,j}/x_{\rm max}$ for all phages.
At short times, the immunity has not built up
yet, and the condition $i \ne k, j \ne k$ is irrelevant.
The only difference between the non-linear and constant growth rates at short
time is a slightly slower increase of the phage population in the
non-linear case.
 The blue and black curves in Fig.\
\ref{fig:population} show phage populations with constant and
density-dependent
growth rates differ only at short times.

In general, we are interested in the case where the phage and
bacteria populations reach steady-state.  In this case, there will
be an effective growth rate of the phage.  This effective growth
rate is $r_0$ in the linear model.  Because the bacteria reach the
maximum population size quickly, and because the non-linear growth
model is different from the linear model only when the bacteria
are below the maximum population size, the growth dynamics of the
non-linear and linear model differ only at short times.  At long
times, most of the bacteria population will still be available to
any given phage for growth, and so $\langle r_k \rangle \approx r
\approx r_0$. In particular, we find $\langle r_k \rangle / r$ is
unity for $t< 200$, and rises only to 0.97 at $t=2000$ for the
parameters we use in section 3.2. Even for the parameters
corresponding to a more effective immune system in section 3.4,
this quantity is unity for $t< 200$, 0.93 for $t=600$, and 0.82
for $t=2000$. These results justify the assumption that most of
the bacteria population is available to growth of any given phage.
Thus, the non-linear and linear growth models only differ at very
short times when the bacterial population is not yet the maximum
size, or at rather long times if the diversity of the phage
population is driven to low values.  The non-linear and linear
growth models would also differ if the bacterial population were
driven extinct, a situation we do not consider in the present
work.

\subsection{Method validation}

\begin{figure}
\centering
\includegraphics[width=100mm]{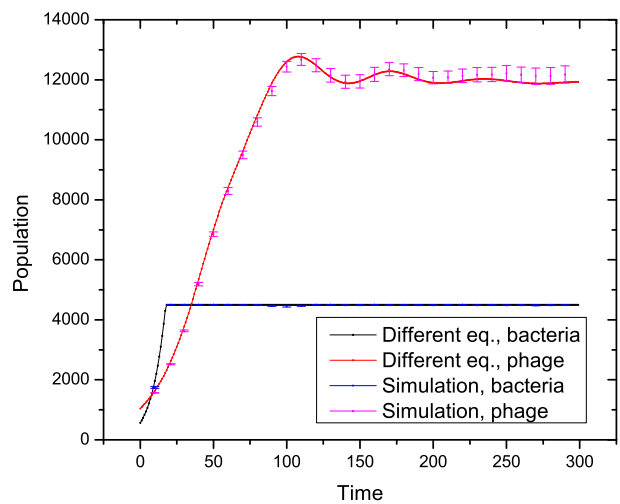}
\caption{The solution of the differential equation and the result of the
stochastic method. Parameters are $ c = 0.15$, $r = 0.045$, $\beta = 2\times 10^{-5}$,
$\gamma = 0.1$, $v_{\rm max} = 17500$, and $x_{\rm max} = 4500$. There are 2 spacers in a CRISPR. The error bars are one standard error.  The bacterial growth rate sets the explicit time scale in this model.
\label{fig:method_population}}
\end{figure}

To validate the stochastic method, we compare it to the solution
of the differential equation. Both results converge to the same
result in the limit of large population. This convergence is evident
in the population versus time curve, Fig.\
\ref{fig:method_population}.

\subsection{Diversity measurements}

The Shannon entropy of spacers at a specific location is a measure of the
diversity of the spacers at that location. A larger Shannon entropy
indicates more diversity of spacers. We, thus, measure the diversity
of spacers by the Shannon entropy: The diversity
for the $i^{th}$ spacer is defined as
\begin{eqnarray}
D_i = - \sum_k p_i(k)\ln p_i(k)
\label{Shannon}
\end{eqnarray}
where $p_i(k)$ is the observed probability to have sequence $k$ at position $i$.

Because new spacers are always added to the leader-proximal end,
the spacer with smaller index is ``younger" than the spacer with
larger index. If the phage do not impose selection pressure on the
bacteria, all spacers are randomly selected and inserted into the
leader-proximal end of the CRISPR, and we will observe homogeneous
diversity at all positions of CRISPR. With selection pressure, the
diversity of CRISPR may decline toward the leader-distal end of
the CRISPR if the distribution of phage genotypes is biased, as
has been observed in experiments
\cite{HorvathJB2008,BarrangouScience2007,WhitakerPLosOne2010,AnderssonScience2008,TysonEM2007}.
Gaps and insertions in the CRISPR array may result in different
bacteria having nearly the same spacer content, but at slightly
shifted spacer positions.  This dephasing will be observed as a
reduction of $D_i$ values relative to what could be observed with
multiply aligned sequences.

We also define a diversity that averages out the
effect of spacer position.  The definition is
\begin{eqnarray}
D = - \sum_k \left[\sum_i p_i(k)/N \right] \ln \left[\sum_i p_i(k) / N\right]
\label{Shannon2}
\end{eqnarray}
where $N$ is the number of spacers within the CRISPR.

In addition, we define the diversity of the phage.  This is
simply Eq.\ (\ref{Shannon}), but applied to the single
protospacer in each phage, rather than spacer $i$ of each
bacteria.

\subsection{Spacer effectiveness}

We count the number of matches between the spacer at position $i$
of the CRISPR and the current phage strains. We use this count as
a measure of protection offered by spacer. The bigger this count
is, the more frequently the spacer can be used, and the
more effective  the spacer is at protecting the
bacteria from the phages. A higher frequency of usage
indicates a strong protection. Since the spacers at the
leader-proximal end are recently acquired and reflect the current
viral environment, these spacers should be highly used and offer
the strongest protection against current phages. Within one
CRISPR, we expect a decline of the protection with respect to
position from the leader-proximal.

\subsection{Recombination in the Phages}
\begin{figure}
\centering
\includegraphics[width=100mm]{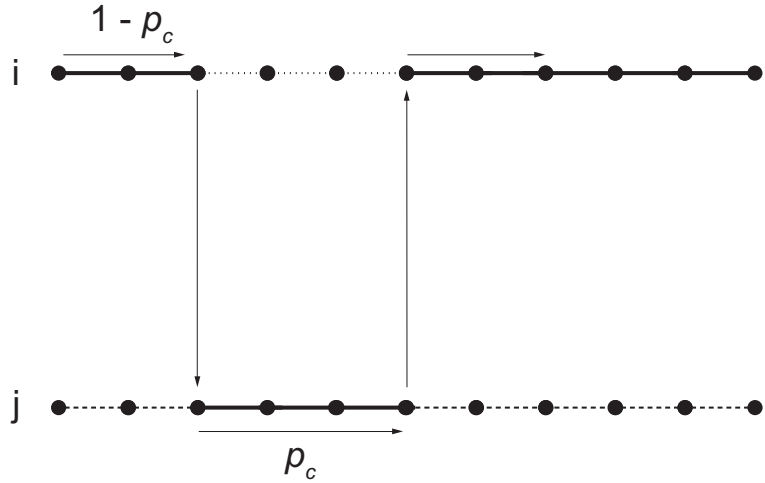}
\caption{When multiple phages infect the same
bacteria, two parents may produce a descendant by the polymerase
copying along one strand with probability $1-p_c$ and switching to
another strand with probability $p_c$. This process
leads to recombination
between the phage genomes. \label{fig:recombination}}
\end{figure}

Another CRISPR-evading strategy of phage is recombination.
Recombination can recombine multiple existing point mutations or
even different strains. When two phages infecting the same cell
recombine, they swap a portion of their genetic materials. This
swapping is a random process. For a given sequence, recombination
happens with a randomly chosen other sequence at a given
probability $\nu$ per sequence per replication. In this way,
mutation rates and recombination rates have the same units. Also
with this definition, equal recombination and mutation rates imply
an equal probability of changing a given sequence by an evolution
event, except for the relatively rare occurrence of recombination
between two identical sequences. We simulate this random process
as follows: To assemble the
 recombined phage from two existing parental strains,
we first pick randomly which sequence to start copying,
then a polymerase continues on that sequence with probability $1-p_c$ or
switches to the other with probability $p_c$ until an entire offspring sequence is created.
See Fig.\ \ref{fig:recombination}.

\section{Results}

% the following is comments
\iffalse
Before 40 seconds, method one and method two show
different results. That means the dependence of bacteria
population is important when investigating the population dynamic
of the phage on the short time scale. After 40 second, method one
and method two show exactly the same result. That means
\fi
%end of the comments

We are interested in the coexistence of bacteria and
phage at long times. In these models, both the total phage and bacteria population grow to their
maximum carrying capacity at long times. Density-dependent and constant growth rates produce the same results.

\subsection{Diversity versus position}

\begin{figure}
\centering
\includegraphics[width=100mm]{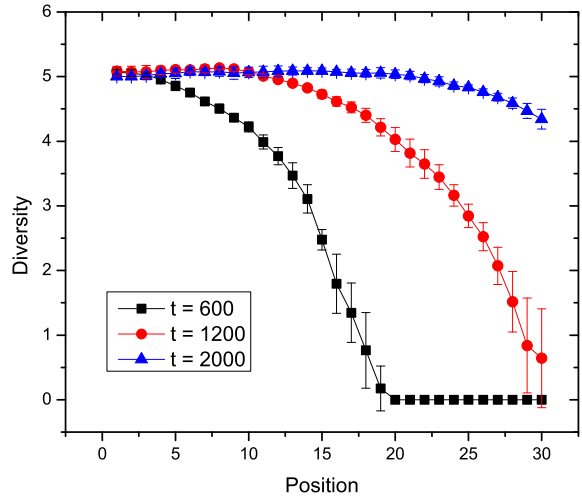}
\caption{Diversity of spacers at CRISPR position
$i$ at different times.
The parameters are the same as in Fig.\ \ref{fig:population}.
 \label{fig:diversityVsposition}}
\end{figure}

The diversity of spacers in the CRISPR system is measured using
Shannon entropy, Eq.\ \ref{Shannon}. We keep track of spacer
diversity with respect to the position of the spacer. This is
shown in Fig.\ \ref{fig:diversityVsposition}. The diversity of the
spacers at the leader-proximal end is higher than the diversity of
the spacer at the leader-distal end.

\subsection{Protection versus position}

\begin{figure}
\centering
\includegraphics[width=100mm]{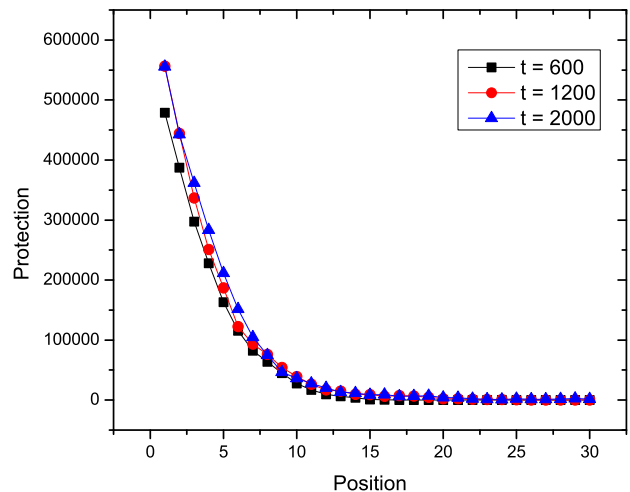}
\caption{Protection afforded by spacers at different
positions of CRISPR at different times. Protection is
defined as the number matches
between the spacer and the protospacers in the
current phage population. The parameters are the
same as in Fig.\ \ref{fig:population}.
\label{fig:protectionVsposition}}
\end{figure}

We define protection as a measure of spacer effectiveness, \emph{i.e}. match of
CRISPR spacers to phages. We calculate the ability of spacers at
position $i$ to protect against the current viral
population. This is shown in Fig.\ \ref{fig:protectionVsposition}.
Since bacteria have the ability to acquire new protospacers
from the phage population, and the insertion of
new spacers happens at the leader-proximal end of the CRISPR, it is
expected that the spacers at the leader-proximal end have the highest
frequency of usage. The protection of the
spacer falls off rapidly with distance from leader. Nonetheless, due to the random loss of spacers,
some infection memory can be lost as time elapses.

\iffalse
\begin{figure}
\centering
\includegraphics[width=100mm]{protectionrise.eps}
\caption{Protection arise a little bit near the leader-distal end after a long time.}
\label{fig:protectionrise}}
\end{figure}
\fi

\subsection{Deletion mechanism}

\begin{figure}
\centering
\includegraphics[width=2.5in]{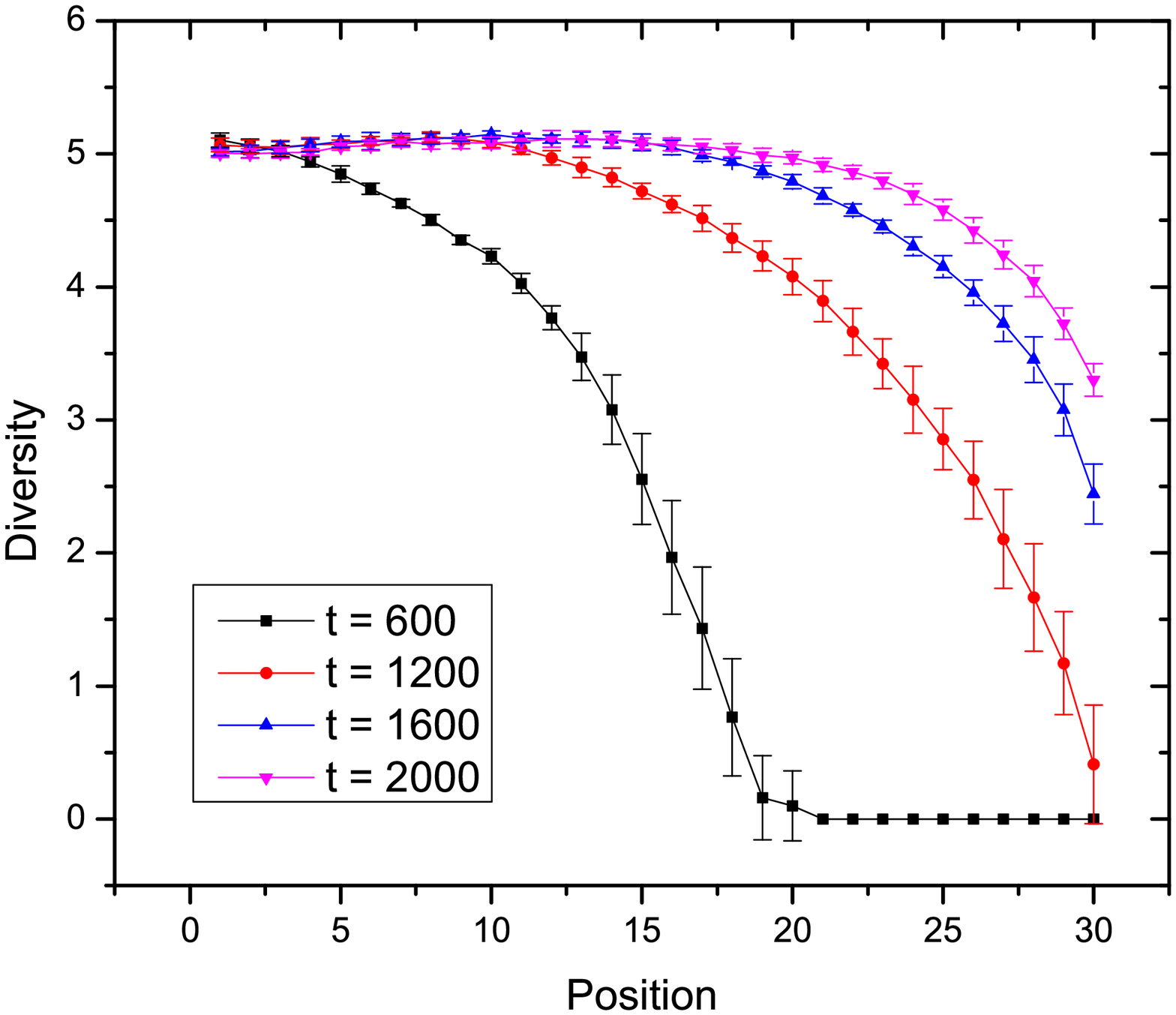}
\includegraphics[width=2.5in]{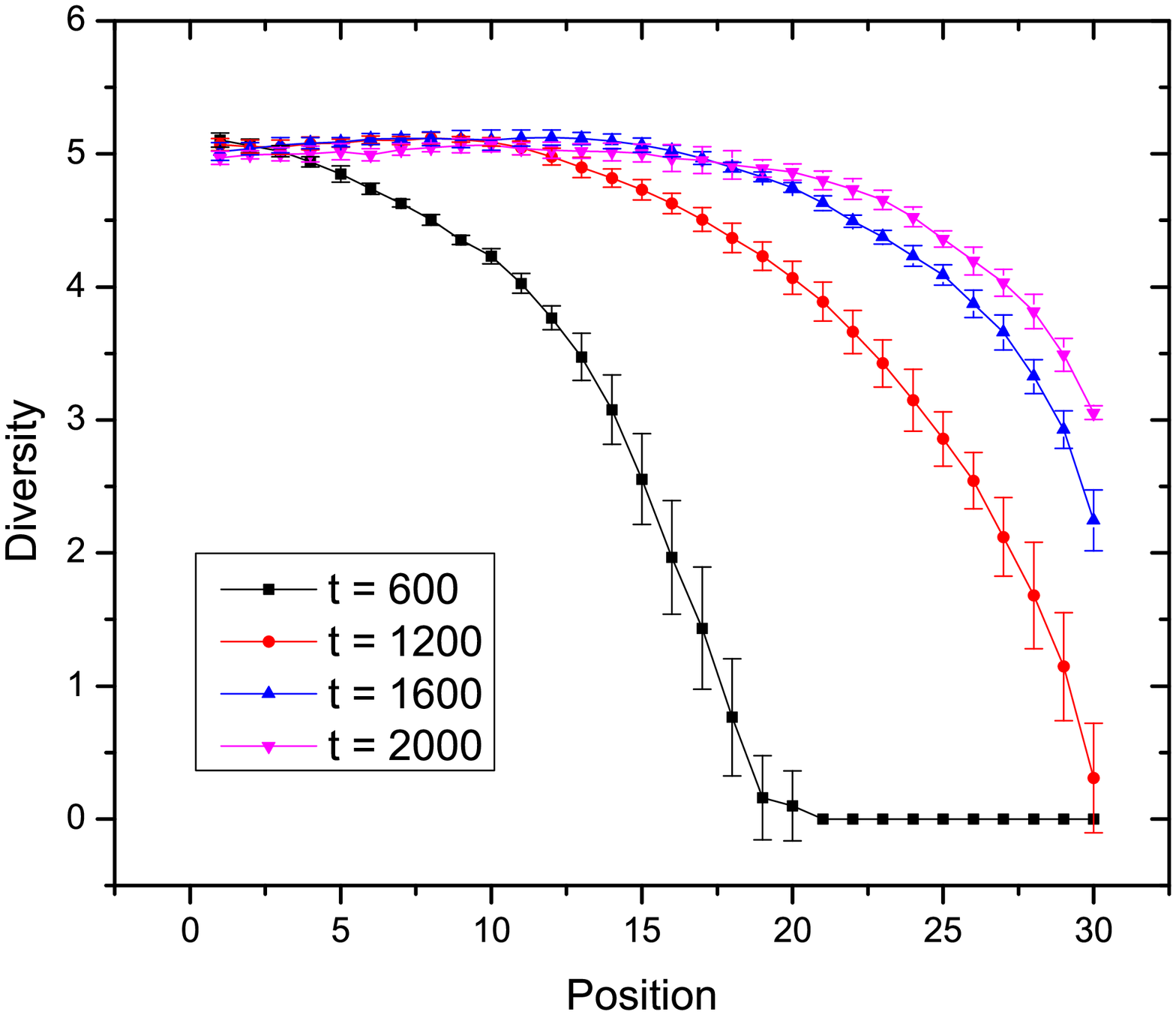}
\caption{Spacer diversity versus location when the
probability of deleting spacer $i$ is proportional to $i$.
a) When the
number of spacers in the CRISPR array is over 30, one spacer is
selected to be deleted with a possibility proportional to its
distance to the leader proximal end.
b) When the number of spacers in the CRISPR array is over
30, one spacer at a random location is deleted.
The parameters are the same
as in Fig.\ \ref{fig:population}. \label{fig:diversitylinear}
%\label{fig:diversityuniform}
}
\end{figure}

The diversity of spacers with respect to the location of spacers for
three different deletion mechanisms is shown in Figs.\
\ref{fig:diversityVsposition} and
\ref{fig:diversitylinear}.
There is a small but significant difference in the dynamics of these
three models.
Diversity versus time shows the same trend for all three methods:
the leader-proximal end is more diverse, and leader-distal end is less
diverse. Although the diversity decreases toward the leader-distal
end of the CRISPR, it decreases the least when the oldest spacer is deleted.
In random deletion, every
spacer in CRISPR has the same possibility to be deleted. Even the newer
spacer, closer to the leader-proximal end, can be
deleted. As a result, the decrease of the diversity at
the leader-distal end from the leader-proximal end is the largest
among the three for random deletion. The decrease of diversity for the linear
deletion mechanism is midway between that for the other two deletion mechanisms
because leader-distal spacers
with less diversity have more possibility to be
deleted.

%\begin{figure}
%\centering
%\includegraphics[width=100mm]{doldest.eps}
%\caption{Spacer diversity versus location when the
%oldest spacer is deleted. When the number of spacers in the CRISPR
%array is over 30, the oldest spacer is deleted from the leader distal
%end. The parameters as the same as in Fig.\
%\ref{fig:population}. \label{fig:diversityoldest}}
%\end{figure}

\subsection{Recombination versus mutation}

We compare the impact of mutation and recombination on phage
evolution. We define the minimum number of mismatches between the
CRISPR and phage required for the phage to escape recognition as
$l$. The bigger the value of $l$, the harder it is for the phage
to escape from the targeting spacer. A value of $l=1$ means if
there is one or greater mismatch between the spacer and the phage
genome, the spacer provides no protection against the phage. A
value of $l=2$ means that CRISPR recognizes the phage even if
there is one mismatch between the spacer and the phage genome. For $l=2$, if
the spacer and the phage differ at one position, the phage is still
recognized and neutralized by the targeting spacer, \emph{i.e.}\ the
CRISPR is more effective. This internal error tolerance makes it
harder for phage to escape by mutating one bit of their
protospacer for $l=2$. If the number of mismatches is greater than one, the
spacer provides no protection against the phage for $l=2$. We show that
there is little difference in the results for point mutation and
recombination when $l=1$. However, when $l=2$, the difference in results
between the point mutation and recombination becomes apparent.
It is widely assumed that $l=1$ describes phage recognition \cite{BarrangouScience2007}.
It seems likely, however, that a protospacer with a single mismatch would also be
recognized, i.e. $l=2$ should apply in at least some cases, and some evidence for single-mismatch recognition has
been observed \cite{DeveauJB2008,MarraffiniScience2008}.

% these are the added sentences. 11/25/2012

\begin{figure}
\centering
\includegraphics[width=2.5in]{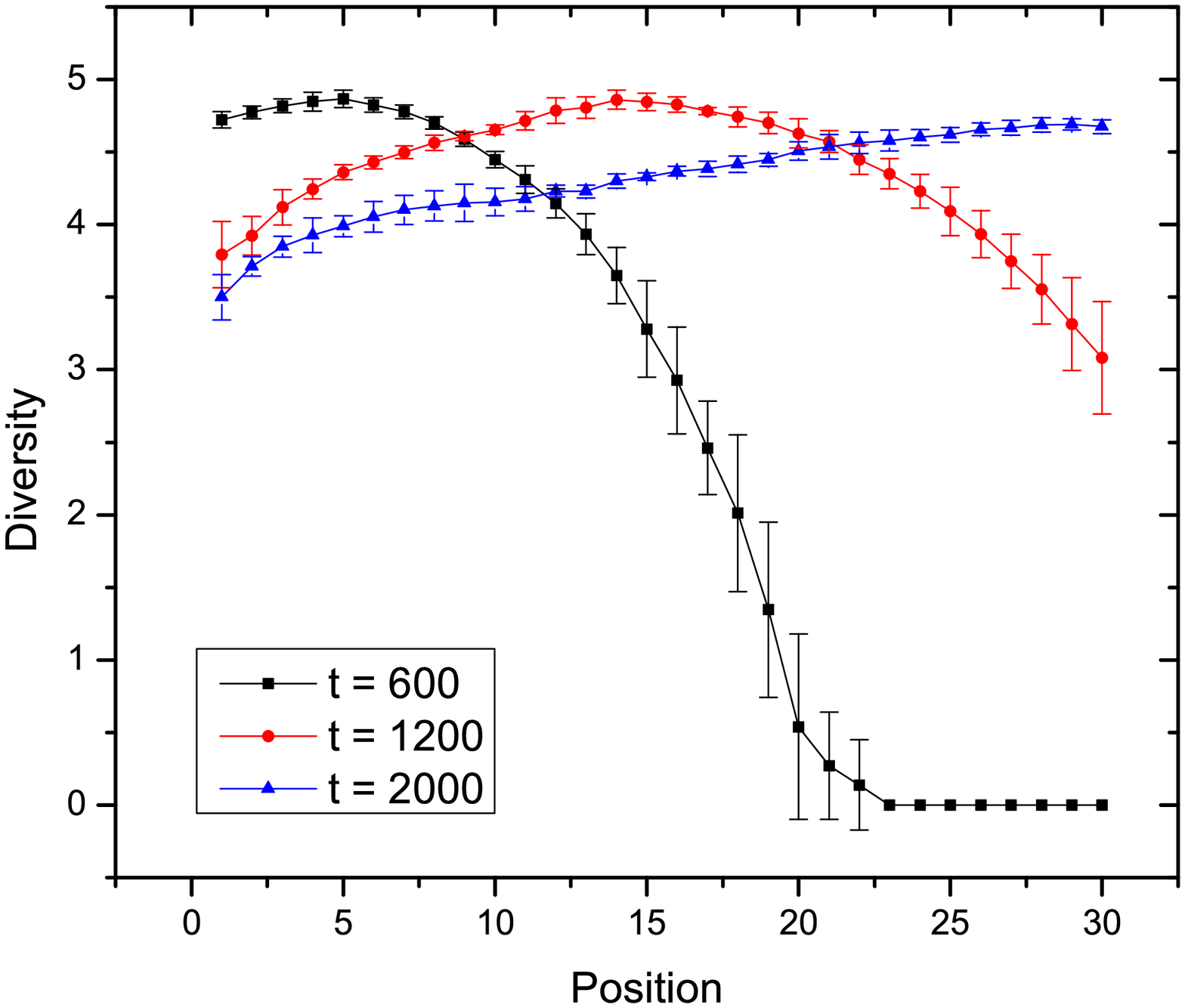}
\includegraphics[width=2.5in]{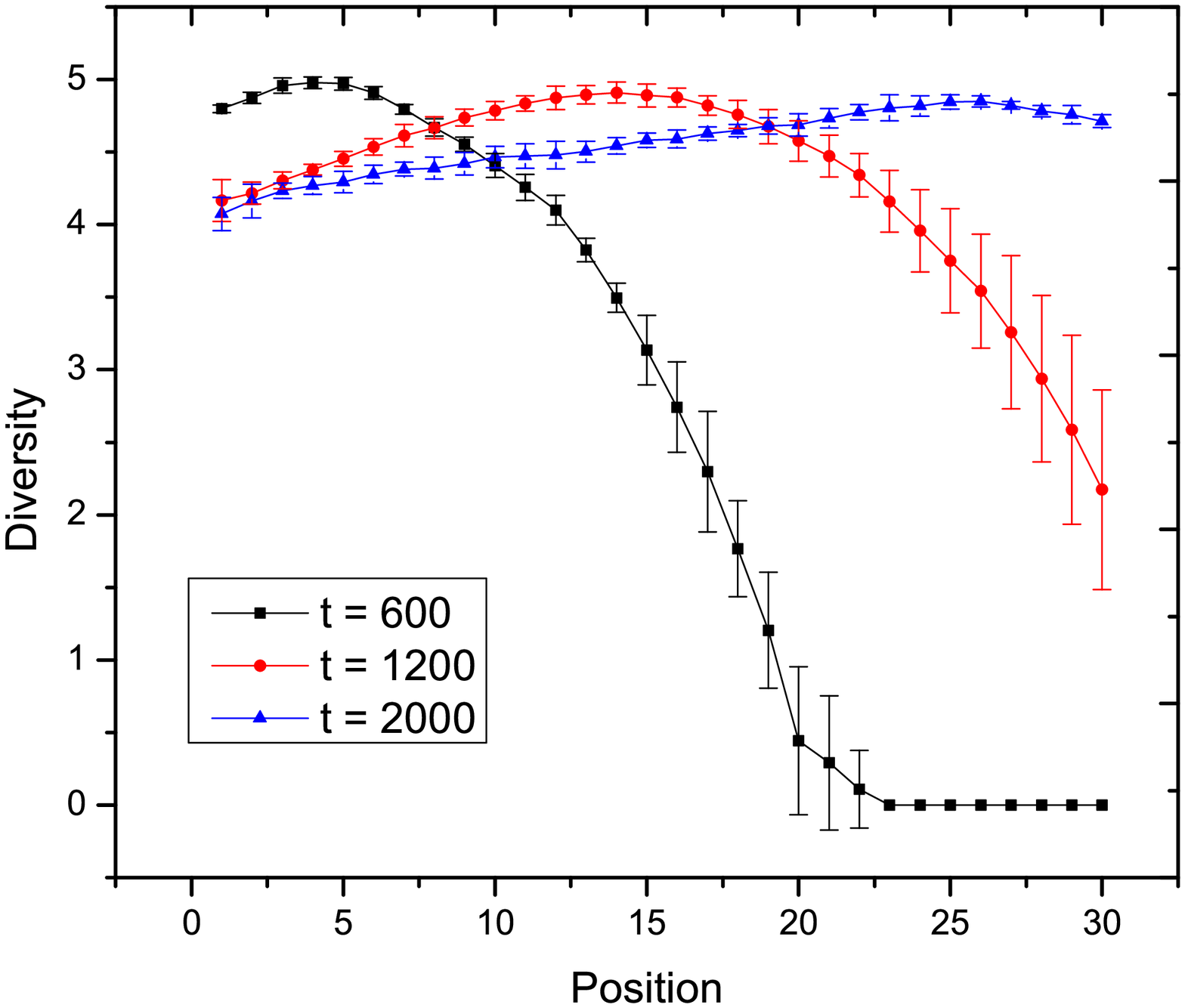}
\caption{Diversity at different positions of CRISPR at different
times for $l=2$ with a) mutation only and b) recombination only.
In this case, CRISPR recognize phage with zero or one mismatch
between the spacer in the bacterium and protospacer in the phage.
The recombination rate per sequence per replication is $\nu=0.01$,
and $p_c = 0.5$. The other parameters are the same as in Fig.\
\ref{fig:population}.
 Spacer diversity is not particularly sensitive to whether the
phage evolve by mutation or recombination.
\label{fig:diversitymutationl2}
%\label{fig:diversityrecombinationl2}
}
\end{figure}

The different CRISPR-evading strategies of recombination and mutation have
minimal impact on the spacer diversity with respect to
position, as shown by Fig.\ \ref{fig:diversitymutationl2}
in comparison to Fig.\ \ref{fig:diversityVsposition}.
For $l=2$, the spacers are
slightly more diverse when phage recombine than they when
mutate. Although recombination allows phage to make a more diverse set of
descendants than does point mutation, the observed effect in the
diversity of CRISPR
is small.  Thus, spacer diversity is not a sensitive measure
to distinguish different CRISPR-evading strategies.

\begin{figure}
\centering
\includegraphics[width=100mm]{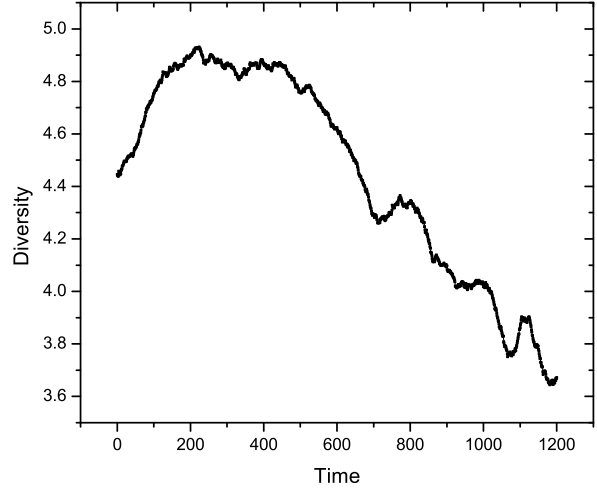}
\caption{Diversity of the phage
for $l=2$ with mutation only.
The parameters are the same as in Fig.\ \ref{fig:diversitymutationl2}.
\label{fig:diversityphage}}
\end{figure}

At long times, the diversity of the leader-proximal spacers decreases.
This is because the diversity of the phage population itself
decreases for large time.  This diversity profile is shown in
Fig.\ \ref{fig:diversityphage}.

We define ``immunity" as a measure of the possibility that
CRISPR will kill phage:
$\beta\sum_k\sum_{i,j}x_{i,j}v_k(\delta_{i,k}+\delta_{j,k})$. The
higher the immunity is, the higher protection the spacer provides.
Figure \ref{fig:k1andk2} shows that recombination
gives phage more chance to survive and the CRISPR
immunity is lower. When $l = 1$,
immunity is similar whether pages escape by mutation or
recombination,
because the
effectiveness of spacer is equal in regard to escape by point mutation or
recombination. When $l = 2$, the immunity is higher against escape by mutation
than it is against escape by recombination. Immunity decays more quickly with
recombination rate
 than with mutation rate, Fig.\ \ref{fig:k1andk2}.

\begin{figure}
\centering
\includegraphics[width=2.5in]{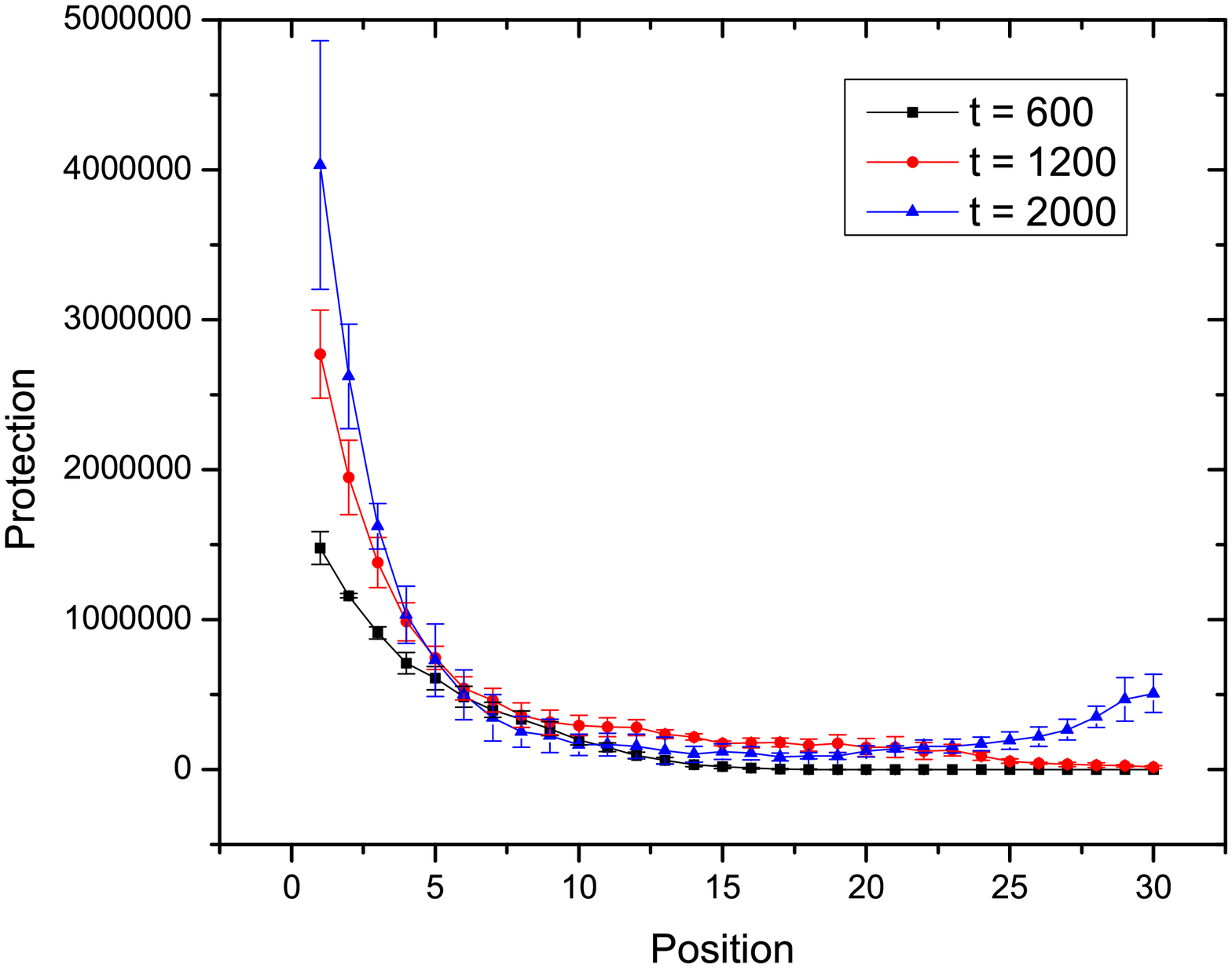}
\includegraphics[width=2.5in]{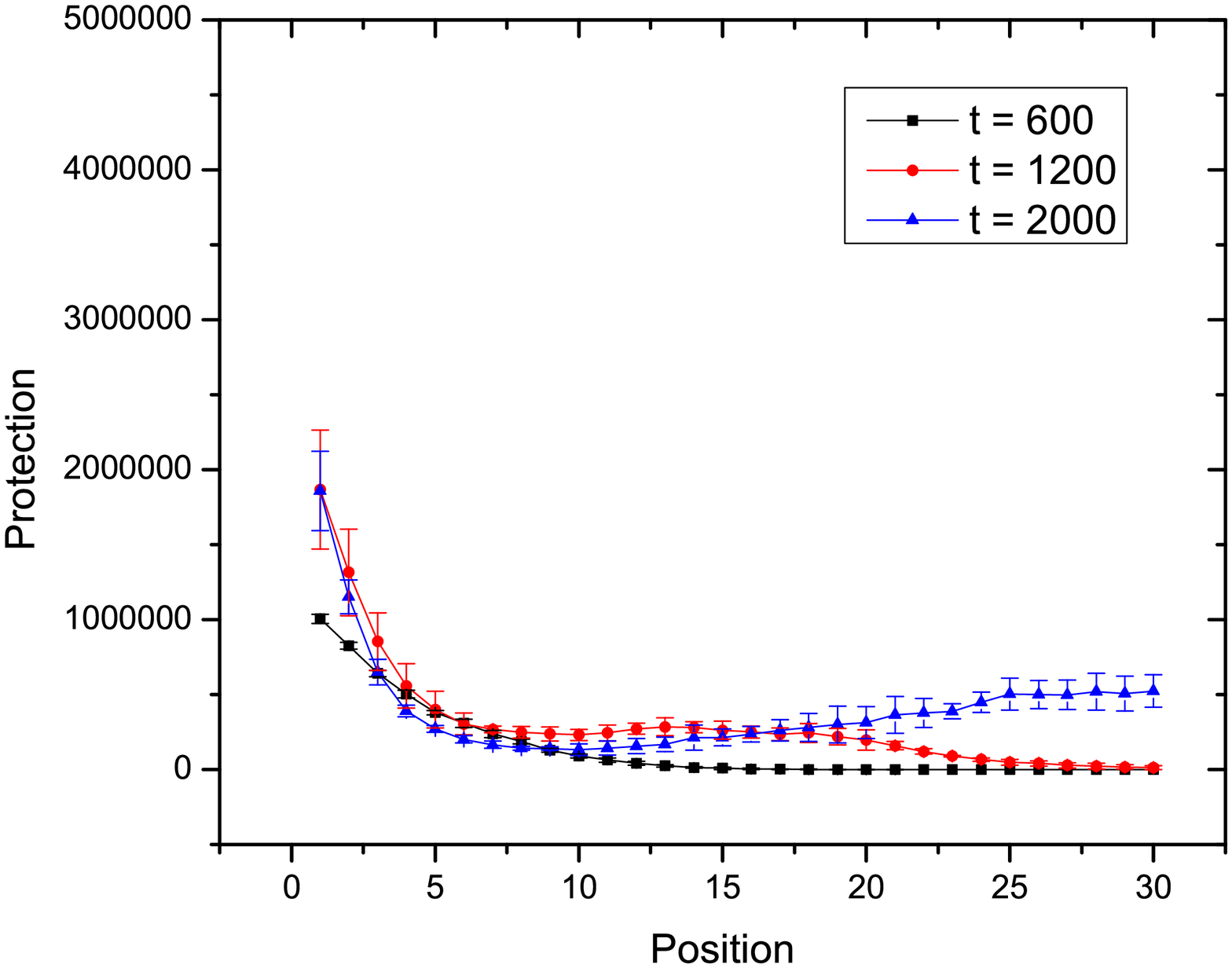}
\caption{Protection at different positions of
CRISPR at different times for $l=2$ with a) mutation only
and b) recombination only. The
parameters are as in Fig.\ \ref{fig:diversitymutationl2}.
\label{fig:Protectionmutationl2}
%\label{fig:Protectionrecombinationl2}
}
\end{figure}

\begin{figure}[ht]
\centering
\subfigure[]{
\includegraphics[scale=0.25] {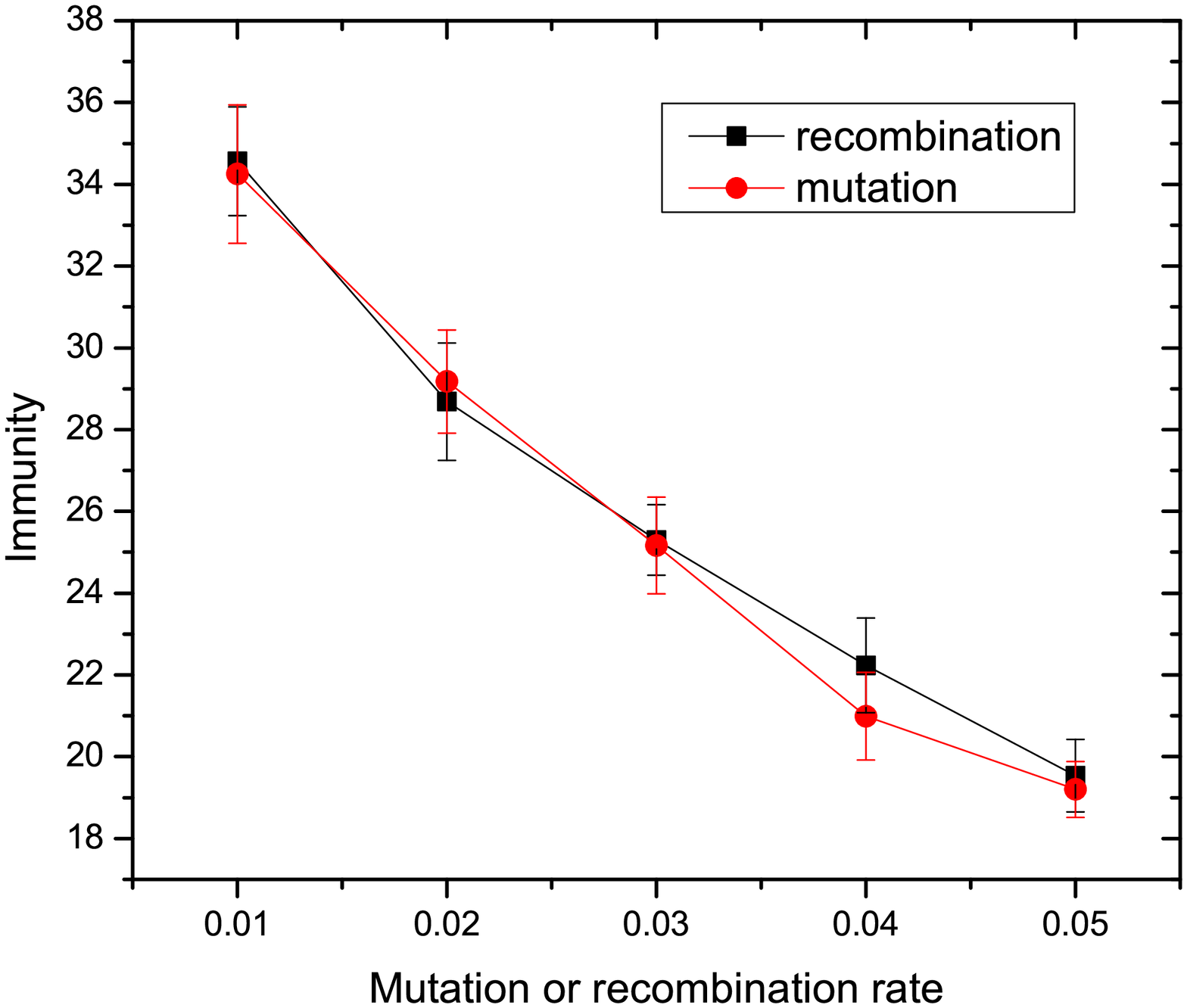}
    \label{fig:k1}
}
\subfigure[]{
\includegraphics[scale=0.25] {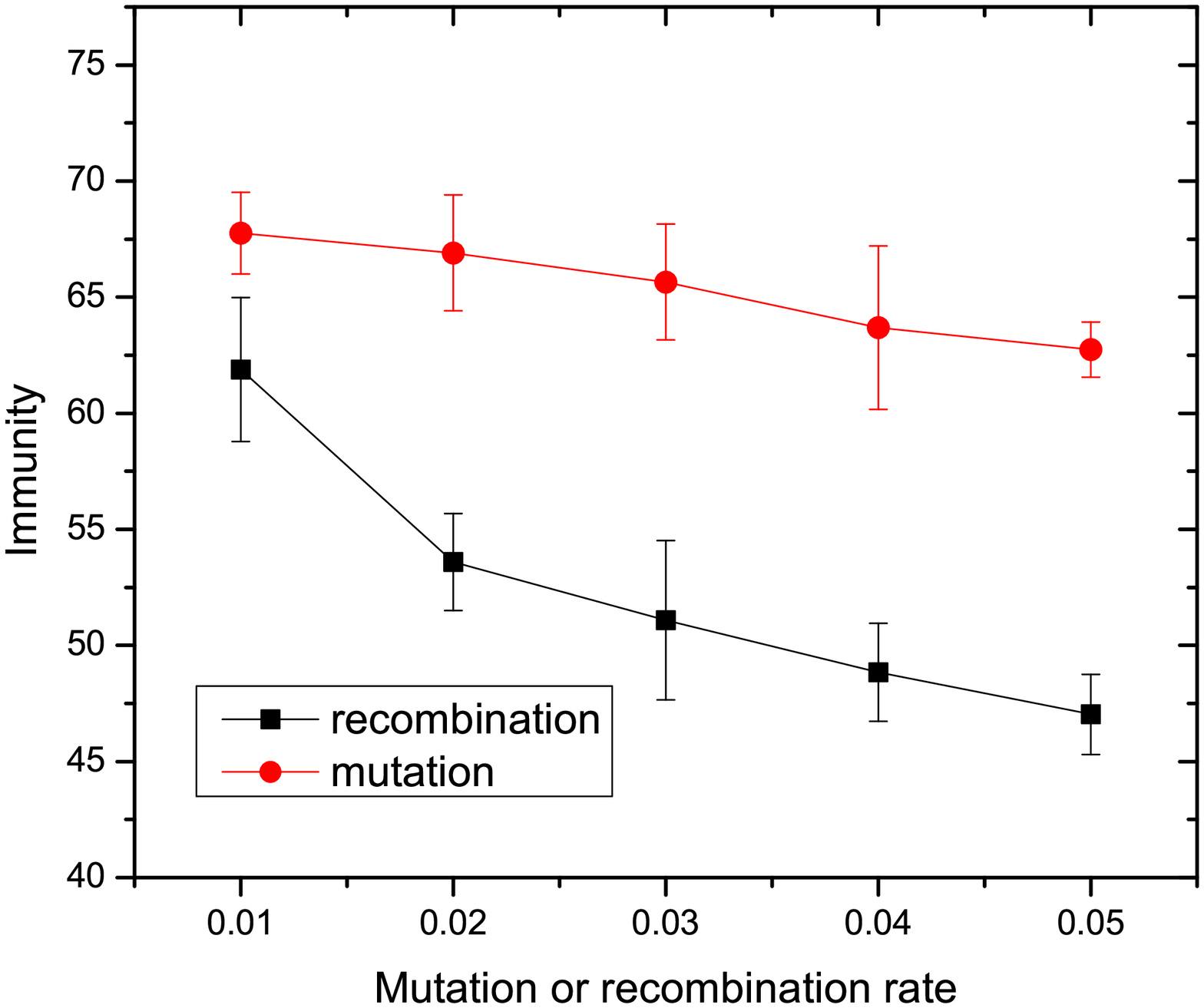}
    \label{fig:k2}
} \caption{Immunity,
$\beta\sum_k\sum_{i,j}x_{i,j}v_k(\delta_{i,k}+\delta_{j,k})$, at
different mutation rate and recombination rate when $l=1$
\subref{fig:k1} and $l=2$ \subref{fig:k2}. The other parameters are as in Fig \ref{fig:population}.
The immunity is averaged over the time range $t=100$ to $t=300$.} \label{fig:k1andk2}
\end{figure}

\iffalse
\begin{figure}
\centering
\includegraphics[width=100mm]{k1mutation.eps}
\caption{Immunity versus mutation rate for $l = 1$. \label{fig:k1mutation}}
\end{figure}

\begin{figure}
\centering
\includegraphics[width=100mm]{k1recombination.eps}
\caption{Immunity versus recombination rate for $l=1$.
\label{fig:k1recombination}}
\end{figure}
\fi
\iffalse
\begin{figure}
\centering
\includegraphics[width=100mm]{k1andk2.eps}
\caption{Immunity versus mutation and recombination rate for $l = 1$ and $ l = 2$. \label{fig:k1andk2}}
\end{figure}
\fi
\iffalse
\begin{figure}
\centering
\includegraphics[width=100mm]{k2mutation.eps}
\caption{Immunity versus mutation rate for $l=2$. \label{fig:k2mutation}}
\end{figure}

\begin{figure}
\centering
\includegraphics[width=100mm]{k2recombination.eps}
\caption{Immunity versus recombination rate for $l=2$.
\label{fig:k2recombination}}
\end{figure}
\fi

\section{Discussion}

We have addressed whether or not
bacteria population dependence should be included in the phage growth rate, $r$.
We have shown that a natural form of nonlinear growth dynamics makes no
difference at long time in the regime where phage and bacteria coexist, although there is a slight
difference at short time. Since we enforce co-existence, this
detail is inessential under the conditions of our study.
 There are multiple strains of phage
and bacteria, and most strains of the phage can
grow in nearly all strains of the bacteria in our simulations.

 The diversity of the spacers at the leader-proximal end
shown in Fig.\ \ref{fig:diversityVsposition}
 is higher than the diversity of
the spacer at the leader-distal end. This result is consistent with
experimental observations on different bacteria
\cite{HorvathJB2008,BarrangouScience2007,WhitakerPLosOne2010,AnderssonScience2008,TysonEM2007}.
The difference in diversity between these two ends decreases as
time elapses as the spacers fill in the CRISPR and the phage
strains randomize due to mutation. This result shows the diversity
of the spacers increases as the diversity of phage increases. This result
is also observed in a more complex microbial community
experimentally \cite{HeidelbergBhayaPlos2009}.

 It has often been assumed that when the CRISPR is
``full'' and spacers are to be deleted, the oldest spacer is
deleted, or the oldest spacer is more likely to be deleted. Not
all mechanisms for spacer deletion are capable of such a biased
removal \cite{BhayaARG2001}. An equal deletion probability for all
spacers is a simpler and perhaps more biologically motivated
assumption.  We have shown that such a uniform deletion
probability does give a spacer diversity which decreases with
distance from leader sequence, in accord with observation. A
uniform deletion rate may be a simple, yet representative model
for spacer deletion.

We have quantified the impact of mutation and recombination on
phage escape from CRISPR recognition.  Mutation and recombination
both allow phage to escape. So far, most theories have assumed
that phages evolve only by point mutation. Here
we have examined the effects of
recombination on the coevolution process,
complimenting previous theoretical studies
\cite{WeitzEvolution2012,HaerterSneppenJBI2011,HaerterSneppenMBIO2012,
SneppenaKrishnabPNAS2012,WeinbergerPLosCB2012,LevinPlos2010}. Data
suggest that recombination is a significant driver of evolution
\cite{AnderssonScience2008}. To quantify the effectiveness of
mutation versus recombination in phage escape, we defined a new
quantity, ``immunity," the rate at which bacteria kill phages.
This immunity is a good measure of the effectiveness of phage
escape. By computing immunity, we quantified and compared the
relative efficiencies of mutation and recombination for phage
escape. There may be selective constraints on what mutations can
occur in the viral protospacer. Consequently, phage need to find
``viable" mutations. Recombination in the phages can combine
beneficial or viable mutations. Furthermore, one mutation may not
necessarily be enough to escape the host CRISPR immunity system,
and it is possible that greater than one mutations is needed in
order for a phage to escape. For both of these reasons,
recombination allows phage to escape CRISPR recognition more
effectively than does mutation alone.

Differing immune pressures become distinguishable in the diversity
measurements at long times. At short times, the diversity results for $l=1$ in Figure \ref{fig:diversityVsposition} and for $l=2$
in Figure  \ref{fig:diversitymutationl2} are similar.
The results differ at longer
times, $t \ge 1200$, in these two figures.

Interestingly, the leader-proximal spacers are less diverse
in Fig.\
\ref{fig:diversitymutationl2}
for $l=2$ than they are for $l=1$. A lower diversity of these spacers
is also observed for smaller mutation or recombination rates.
When the phage is less able to escape the CRISPR, the diversity
of the phage population decreases at long times.  For this reason, the
diversity of the spacers incorporated at later times, \emph{ i.e.}\ the leader-proximal
spacers, is lower than that of spacers incorporated earlier, \emph{i.e.}\ the
spacers a bit farther from the leader.

If the bacteria are killed less by the phage, for example by having a more effective immune system, they are
able to add a greater number of spacers and to fill up their CRISPR array more quickly.  As the spacers fill in the
leader-distal CRISPR positions, the diversity rises above the initial value of zero.  It is for
this reason that leader-distal diversity as
a function of position for smaller mutation rate, smaller recombination rate, or larger $l$
are above those for higher mutation rate, higher recombination rate, or smaller $l$.
The interplay between the decrease of phage population diversity at long times and the
filling in of the CRISPR array leads to the non-monotonic diversity of
spacers with position in Fig.\ \ref{fig:diversitymutationl2}.
The protection as a function of position can also be non-monotonic, as is
Fig.\ \ref{fig:Protectionmutationl2},
% and \ref{fig:Protectionrecombinationl2},
due to a decreasing diversity of phage with time and the diversity
of leader-distal spacers being greater than that of intermediate spacers.

Protection of CRISPR is a better measure to differentiate the two
CRISPR-evading strategies of mutation or recombination. % this is the end of added sentences
From the figures of protection versus position, we can see that
when $l=2$, the protection of CRISPR is lower when the phage
recombine, Fig.\ \ref{fig:Protectionmutationl2}b, than
mutate, Fig.\ \ref{fig:Protectionmutationl2}a. That is, recombination
allows the phage to escape the CRISPR system more easily. This result
illustrates that recombination is a more efficient CRISPR evading
strategy for phage.

\section{Conclusion}

The CRISPR/Cas system plays a crucial role in bacteria and phage
coevolution. By adding and deleting spacers, bacteria are evolving
dynamically under the selection pressure imposed by phage undergoing
point mutation and recombination.  The stochastic model used in this
work captures the essential features of the
CRISPR/Cas system, giving rise to the fascinating characteristics
coexisting bacteria and phage system.
The rich variety of spacers within the
CRISPR locus captures the history of bacteria and phage coevolution.

As the ``ancient''  winner with better fitness, the leader distal
spacers are more homogeneous than the leader proximal spacers.
This result has previously been observed under a wide range of
model parameters \cite{DeemPRL2010}. Bacteria with more effective
immune systems, or bacteria attacked by phage that mutate more
slowly, have higher fitness and are able to more quickly fill
their CRISPR array with spacers. This result is rather intuitive
and expected to hold under rather general
conditions\cite{WeitzEvolution2012,HaerterSneppenJBI2011,HaerterSneppenMBIO2012,
SneppenaKrishnabPNAS2012,WeinbergerPLosCB2012}.

 Spacer diversity
is not particularly sensitive to whether the phage evolve by
mutation or recombination. This result may be  a bit surprising.
It is understood to be a result of recombination between two
random phage strains almost always leading to a new phage strain,
and, therefore, identical in effect to mutation.

 Different
mechanisms of spacer deletion subtly affect the distribution of
spacers in CRISPR. Random deletion of spacers
\cite{HorvathJB2008,DeveauJB2008} leads to a modestly slower rate
of filling in the CRISPR array than does a mechanism of deleting
only the leader-distal spacer.  This result is because random deletion
removes non-terminal spacers, which inhibits growth.

The protection or immunity that CRISPR confers upon bacteria is
sensitive to the effectiveness of CRISPR-phage recognition,
distinguishing between whether $l=1$ and $l=2$ mismatches are
required for phage to escape recognition.  Protection and immunity
are also sensitive to the mechanism of phage escape, easily
distinguishing different rates of phage evolution.  Recombination
is seen \cite{AnderssonScience2008} to be more effective in
allowing phage to escape CRISPR recognition when greater numbers
of mutations are required for escape, $l=2$. This result is simply
because recombination is most often with a quite different strain,
and so the produced recombinant has more contained variation than 
mutation would provide. It is likely that phage recombination is a
significant generator of phage diversity in the wild.

\section*{Acknowledgments}
This research was partially supported by the US National Institutes of
Health under grant number  1 R01 GM 100468--01.

\appendix
\section{Table of parameters}

\begin{center}
 %   \begin{tabular}{ | l | l | l | p{5cm} |}
    \begin{tabular}{ | l | p{10cm} |}
    \hline
    Parameter & Meaning \\ \hline
    $x_{i,j}$ & The
population of bacteria containing spacers $i$ and $j$   \\ \hline
    $x_{\rm max}$ & The maximum bacteria population  \\ \hline
    $v_k$ & The phage population containing protospacer $k$  \\ \hline
    $v_{\rm max}$ & The maximum phage population \\ \hline
    $c$ & The bacteria growth rate \\  \hline
    $r$ & The phage growth rate \\ \hline
    $\beta$ &  The bacteria exposure rate\\ \hline
    $\gamma$  &  The probability of acquiring a new spacer \\ \hline
    $\mu$  & The mutation rate \\   \hline
    $\nu$ & The recombination rate \\ \hline

    \end{tabular}
\end{center}

\bibliographystyle{unsrt}
\bibliography{CRISPR}

\end{document}